\documentclass[twocolumn,amssymb,amsmath,superscriptaddress,aps,prl,reprint]{revtex4-1}
\usepackage{graphicx}
\usepackage{dcolumn}
\usepackage{bm}
\usepackage{subfigure}

\begin{document}

\title{Spontaneous formation of synchronization clusters in homogenous neuronal
 ensembles induced by noise and interaction delays}

\author{Igor Franovi\' c}
\affiliation{Faculty of Physics, University of Belgrade, PO Box 44, 11001 Belgrade, Serbia}

\author{Kristina Todorovi\' c}
\affiliation{Department of Physics and Mathematics,Faculty of Pharmacy, University of Belgrade,
Vojvode Stepe 450, Belgrade, Serbia}

\author{Neboj\v sa Vasovi\' c}
\affiliation{Department of Applied Mathematics, Faculty of Mining and Geology, University of Belgrade, PO Box 162, Belgrade, Serbia}

\author{Nikola Buri\' c}
\email{buric@ipb.ac.rs}
\affiliation{Scientific Computing Lab., Institute of Physics, University of Beograd, PO Box 68, 11080 Beograd-Zemun, Serbia}%

\date{\today}% It is always \today, today,

\begin{abstract}
Spontaneous formation of clusters of synchronized spiking in a structureless ensemble of equal
stochastically perturbed excitable neurons with delayed coupling is demonstrated for the first time.
The effect is a consequence of a subtle interplay between interaction delays, noise and the excitable
character of a single neuron. Dependence of the cluster properties on the time-lag, noise intensity
and the synaptic strength is investigated.
\end{abstract}


\maketitle

Collective behavior in large ensembles of physiological and inorganic systems can be
reduced to that of coupled oscillators engaged in various synchronization
phenomena.
%Large populations of coupled oscillators engaged in various synchronization
%phenomena present a canonical setup for resolving the collective behavior in a
%myriad of physiological and inorganic systems.
In terms of macroscopic coherent rhythms,
it may either be the case where all the units are recruited into a giant component or the
case of cluster states characterized by exact or in-phase intra-subset and lag inter-subset synchronization.
The spontaneous onset of cluster states is of particular interest to neuroscience \cite{clusters}
for the conjectured role in information encoding, as well as for participating in motor coordination
or accompanying some neurological disorders. The approach to clustering has mostly relied on modeling
neurons as autonomous oscillators, treating separately the question of whether the proposed
mechanisms may be robust under noise \cite{us1} and transmission delays \cite{PitovskyN}. We explore
a new mechanism which rests on the excitable character of neuronal dynamics and mutual adjustment
between noise and time delay to yield the self-organization into functional modules within
an otherwise unstructured network.

For the instantaneous couplings, the research on populations of excitable neurons has
covered pattern formation due to local inhomogeneity \cite{FHN1}, or has invoked a scenario
where noise enacts a control parameter tuning the dynamics of ensemble averages between the three
generic global regimes \cite{FHN2}. Distinct from the layout with complex connection topologies,
here it is demonstrated how coupling delays do alter the latter landscape in a significant fashion,
giving rise to an effect one may dub the cluster forming time-delay-induced coherence resonance.
In part, the strategy to analyze global dynamics rests on deriving the mean-field (MF) approximation
for the exact system. The likely gain from the MF treatment is at least twofold: except for allowing
one to extrapolate what occurs in the thermodynamic limit $N\rightarrow\infty$,
it may serve as an auxiliary means to discriminate between the effects of noise and time delay.
Unexpectedly, the MF model undergoes a global bifurcation at certain parameter values where the exact
system shows an onset of cluster states.

\emph{Network dynamics and the tools to analyze it}-- We focus on an $N$-size population of all-to-all
diffusively coupled Fitzhugh-Nagumo neurons, whose local dynamics is set by
\begin{align}%
\epsilon dx_i&=(x_i-x_i^3/3-y+I)dt+\frac{c}{N}\sum_{j=1}^{N}[x_j(t-\tau)-x_i(t)]dt, \nonumber\\
dy_i&=(x+b)dt + \sqrt{2D}dW_i, \label{eq1}
\end{align}
where the activator variables $x_i$ embody the membrane potentials, while the recovery variables $y_i$
mimic the action of the $K^+$ membrane gating channels. $c$ denotes the synaptic strength and $\tau$
stands for the coupling delay, both parameters for simplicity assumed homogeneous across the ensemble.
The $\sqrt{2D}dW_i$ terms represent stochastic increments of the independent Wiener processes, i.e. the white noise.
 %whose
%expectation values and mutual correlations satisfy $<dW_i>=0$ and $<dW_idW_j>=\delta_{i,j}dt$, respectively.
For the external stimulation holds $I=0$, whereas the small parameter $\epsilon=0.01$ warrants a clear
separation between the fast and slow time scales. Selecting $b=1.05$, the neurons are
poised near the Hopf bifurcation threshold $b=1$, which places them in an excitable regime where each
possesses a single equilibrium. An adequate stimulation, be it by the noise or the interaction term,
may evoke a large excursion of membrane potential, passing through the spiking and refractory states
before it loops back to rest.
%such that it passes through the spiking and refractory states before looping back to rest.

%For isolated neurons, in a range of intermediate $D$ the
%coherence resonance steps in, which is reflected by neurons undergoing limit cycles (LCs) that may
%be broken, in analogy to their deterministic counterparts, into slow motions $O(1)$ along the
%refractory and relaxation branches of the slow manifold and the fast motions $O(\epsilon)$ occurring
%in between.

 To characterize the degree of correlation between the firing events, we use primarily
the interneuron spike train coherence \cite{coherence} $\kappa_{ij}={\sum_{k=1}^{m}X_i(k)X_j(k)}/{\sqrt{\sum_{k=1}^{m}X_i(k)X_j(k)}}.$
This requires one to split the simulation period $T$ into bins $k$ of length $\Delta=T/m$,
awarding each neuron a variable $X_i(k)=1(0)$, contingent on whether a spike was triggered
or not within the given time bin, respectively. As with all the quantities below, we have been
careful to exclude from calculations the transient behavior. The spike threshold and the time
bin are set to $X_0=1$ and $\Delta=0.008$, verifying that no change of the results occurred if
$X_0$ or $\Delta$ were reduced. The distribution of the $\kappa_{ij}$
values may serve to distinguish between the homogeneous and clustered network states.
Another aspect we are interested in is whether the clustered states are monostable or coexistent
with the homogeneous ones at the given network size. To probe this, we have monitored if the
values of the global coherence $\kappa=\frac{1}{N(N-1)}\sum\limits_{i,j=1;i\neq j}^{N}\kappa_{ij}$ for
different realizations at the fixed parameters clumped together, expecting bunching into distinct
groups as evidence of multistable behavior.
%to be a signature of multistable behavior.

Addressing the temporal structure of the network states, it is useful to look into
the distribution of the local neuron jitters $r_i$ \cite{jitter}. They represent the
normalized variations of the interspike intervals $T_k$
%within an $x_i$ time series,
extracted from $x_i(t)$, $r_i=\sqrt{<T_k^2>-<T_k>^2}/<T_k>$, with smaller values indicating
better regularity. The modality and the width of the $r_i$ distribution over
the population may serve as rough indices on how the cluster dynamics is mutually adjusted.
In the final part, we analyze the behavior of the ensemble averages $X=1/N\sum_{i=1}^{N}x_i$ and $Y=1/N\sum_{i=1}^{N}y_i$, where the former increases if a larger fraction of neurons fire in synchrony.
The results for the exact system are compared to those of the approximate MF model \cite{us4}. The
latter presents a two-dimensional set of delayed differential equations
\begin{align}%
\epsilon\frac{dX(t)}{dt}&=X(t)-X(t)^3/3- \nonumber \\
&-\frac{X(t)}{2}\left\{1-c-X(t)^2+\sqrt{[c-1+X(t)^2]^2+4D}\right\} \nonumber\\
&-Y(t)+c[X(t-\tau)-X(t)], \nonumber \\
\frac{dY(t)}{dt}&=X(t)+b, \label{eq4}
\end{align}
derived within a cumulant approach by employing the Gaussian approximation.
%, which has all the cumulants beyond the second order negligible.

We note that the results for the exact system refer to a network of $N=200$ neurons,
applying independently a method from \cite{MG00} to verify no qualitative changes in
the clustering behavior for larger $N$.
%by %the Euler routine with a time step $\Delta t=0.002$ and refer to
%Cluster states have nonetheless been verified to persist if the system size is
%increased.

\begin{figure}[t]%2
%\begin{minipage}[t]{\columnwidth}
\centering
\includegraphics[scale=0.35]{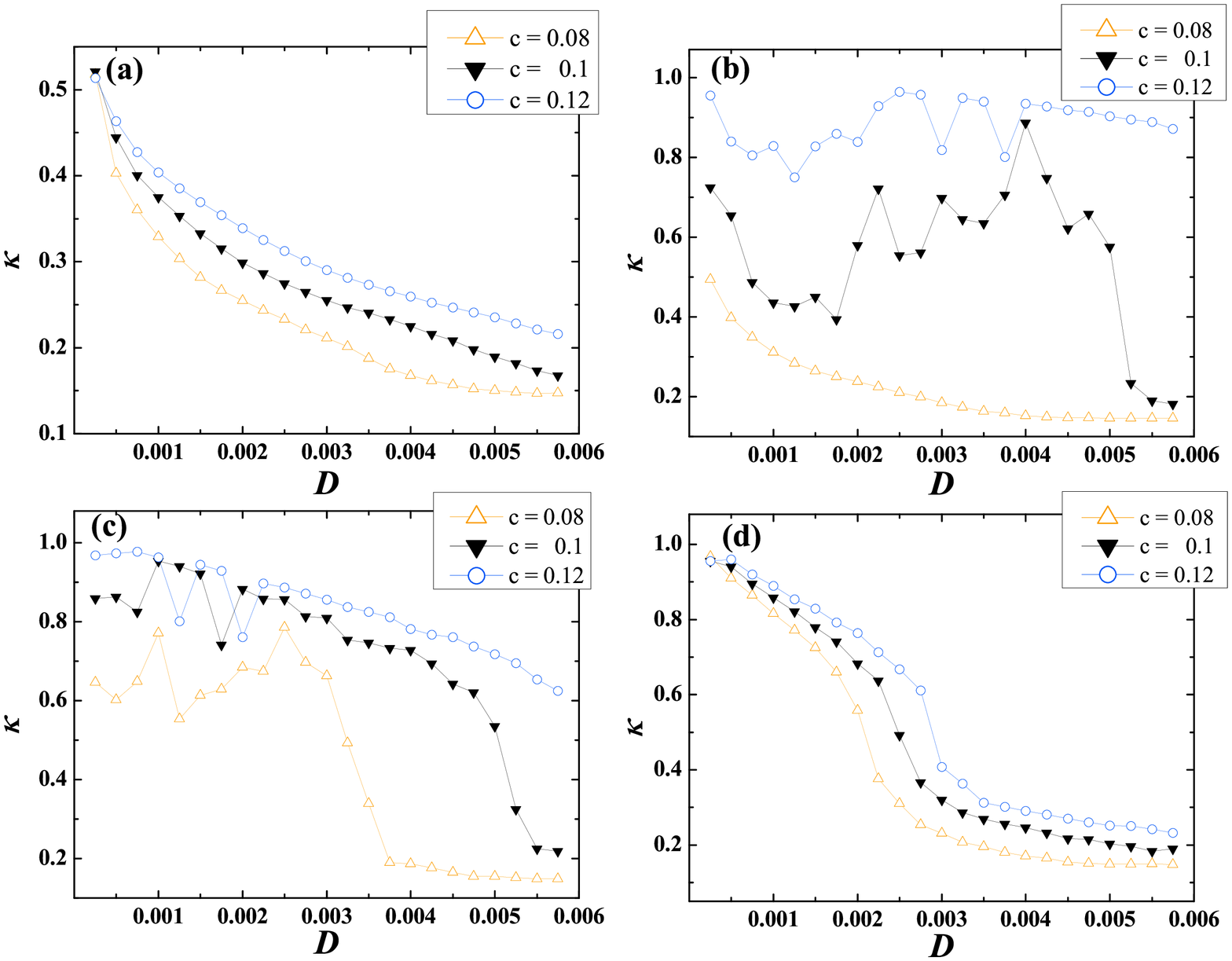}
%\end{minipage}
\caption{(color online). Profiles of the $\kappa(D)$ families of curves over the
synaptic strengths $c=0.08, 0.1$ and $0.12$ display strong dependence on the delay,
increasing from $\tau=2$ in (a), $\tau=6$ in (b), $\tau=10$ in (c) to $\tau=11$ in (d).
%At some instances,
 The location of "wells" may point to the emergence of the clustered
states.  \label{Fig1}}
\vspace{-0.8cm}
\end{figure}

\begin{figure}[t]%2
%\begin{minipage}[t]{\columnwidth}
\centering
\vspace{-0.2cm}
\hspace{-0.4cm}
\includegraphics[scale=0.31]{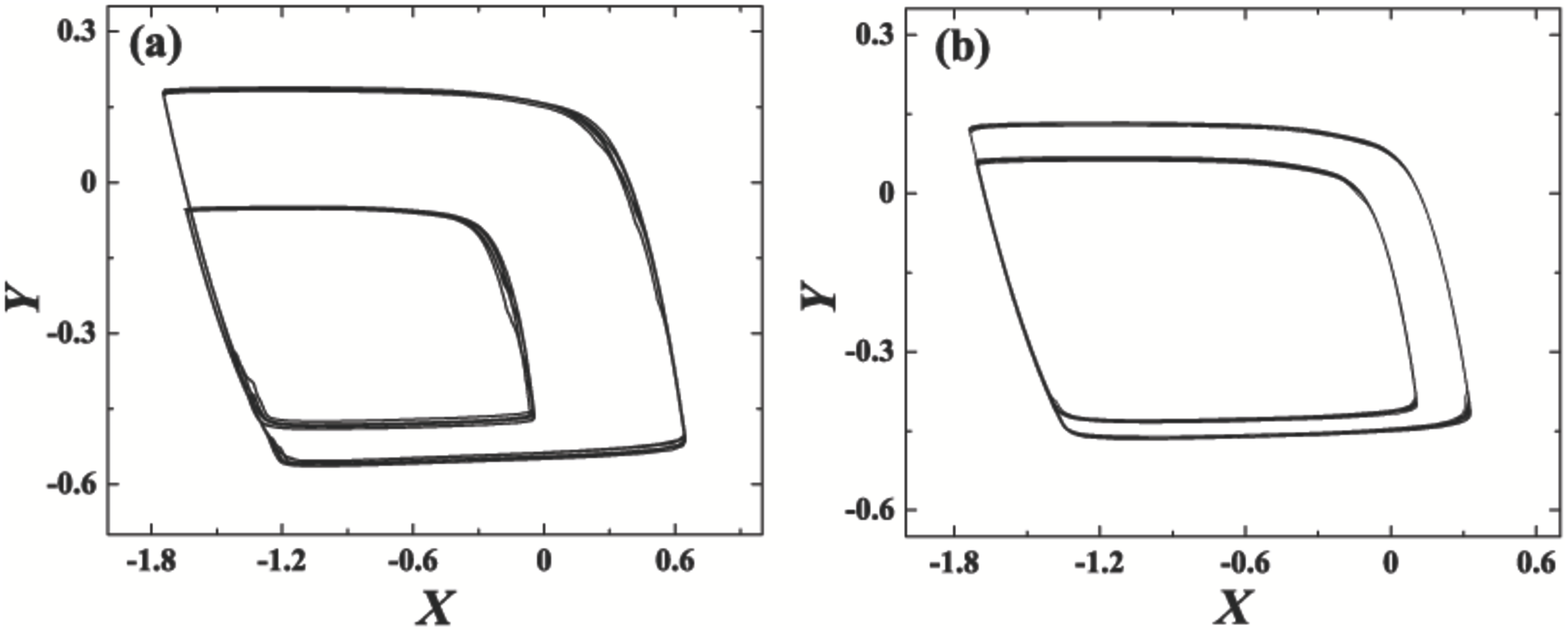}
%\end{minipage}
\caption{Global PPs for the $2$-cluster states show twisted LCs, whereby the two
discernable segments reflect the alternate firing of the neuron subsets.
The $N_1/N_2$ ratio depends on the interplay of $D$ and $\tau$, as seen
from the examples $\tau=2, D=0.00025, c=0.1$ in (a) and $\tau=5, D=0.0005,
c=0.1$ in (b).
%The $\chi(N)$ plot in the inset of (a) implies the near asymptotical behavior is reached for $N\approx200$.
\label{Fig2}}
\vspace{-0.8cm}
\end{figure}

\begin{figure}[t]%2
%\begin{minipage}[t]{\columnwidth}
\vspace{-0.3cm}
\centering
\includegraphics[scale=0.59]{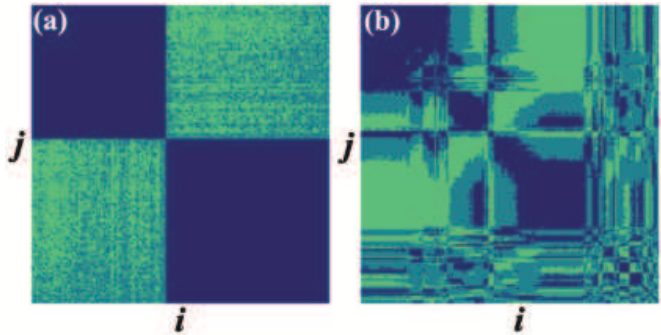}
%\end{minipage}
\caption{(color online). Rearranged coherence matrices for $\tau=5, D=0.0005, c=0.1$ in (a)
and $\tau=10, D=0.0013, c=0.1$ in (b) imply the strong cluster separation in the $2$-cluster states
and mixing between the clusters in the $3$-cluster case. Darker shading reflects higher coherence.
%The higher coherence is represented by darker shading.
\label{Fig3}}
\vspace{-0.8cm}
\end{figure}

\emph{Results}-- To get a sense of what may be the parameter ranges to admit the cluster
states, we plot the $c$--families of the $\kappa$ curves in dependence of $D$ for different
$\tau$. Without the delay, the curves would conform to a stereotype profile, where one
distinguishes between the three "regular" segments for very small, intermediate and large $D$,
showing first a reduced $\kappa$ due to incoherent oscillations, then steady high values for
the coherent ones and the decaying segment at $D$ where the stochastic dynamics prevails.
However, from Fig. \ref{Fig1} we learn how this is upheld for some $\tau$, say $\tau=11$, but
is violated manifestly at the "cluster-resonant" values $\tau=2,6,10$. The "wells" seen at
approximately $D\in(0.001,0.003)$ in Figs. \ref{Fig1}(b) and \ref{Fig1}(c) may occur for just
two reasons, as $\kappa$ decreases either for the incoherent or the clustered states. The latter
alternative is supported by the coherence matrices in Fig. \ref{Fig3}, which are discussed
shortly. The importance of the $D$ - $\tau$ adjustment for the clustering effect is
also witnessed by the $c$-dependence within the families in Fig. \ref{Fig1}: the stronger the
interaction term, the more salient is the picture of "irregularity" sections immersed into
a "regular" curve profile. Increasing the delay, the cluster states first occur, apparently
monostable, around $\tau=2$ for the small $D=0.00025$, whereby the typical phase portrait (PP)
projection shows twisted orbits with two clearly discernable segments, see Fig. \ref{Fig2}(a).
These reflect the two macroscopic fractions of the population firing alternately, such that the
homogeneous network dynamically splits into clusters of mutually synchronized neurons, with the
clusters locked in antiphase. The frequency entrainment is indicated by the
shape of the $r_i$ distribution, which peaks sharply around $<r>_m=0.01$.
%We further find that the $2$-cluster state is stable against increasing the network size.
We tested the invariance of clustering with $N$ via the asymptotical behavior of the
quantity $\chi^2_N=\frac{\sigma_X^2}{\frac{1}{N}\sum_{i=1}^N \sigma_{x_i}^2}$,
where $\sigma_X^2=<X(t)^2>_t-<X(t)>_t^2$ and $\sigma_{x_i}^2=<x_i(t)^2>_t-<x_i(t)>_t^2$
holds. If the cluster states endure, there should be a residual component
$\chi(\infty)\in(0,1)$ in the large $N$ limit \cite{MG00}. For this and the remaining cases,
the onset of such a regime is found around $N\approx200$, implying that no
qualitatively novel phenomena occur above this system size.
An interesting observation is that the cluster configuration $\left\{N_1,N_2\right\}$,
determined by the fractions' sizes, fluctuates around the ratio $2:1$ for different
stochastic realizations and appears to aggregate with enhancing $N$. For certain $\tau$,
the $2$-cluster state also emerges outside the $D$-region
%that may be interpreted as intermediary between the globally incoherent and coherent dynamics.
delimiting the incoherent and coherent global regimes. This holds for $\tau=5$ and $D\in(0.0004,0.0008)$,
%where the firing of clusters is also arranged
where the cluster layout is also such that if one is active, the other remains refractory. The
$r_i$ distribution maintains a narrow form, but its maximum shifts to $<r>_m\approx0.19$. Though one
retrieves the general picture from above, a variance is that larger $\tau$ seems to favor the partition $N_1/N_2\approx1:1$, see PP in Fig. \ref{Fig2}(b). The $1:1$ ratio is preferred both for increasing $N$
and if the delay is set to $\tau=6$.
%at $\tau=5$ and if the delay is enhanced to $\tau=6$ for the same $D$-range.

The clustered states so far may be cast as stationary in the sense of stability against
neurons switching between the clusters. We also report on the existence of $3$-cluster
states that may be considered "dynamical", with the neurons able to jump to and from clusters. Such an outcome arises for the stronger noise $D\approx0.0013$, once the delay is increased to $\tau=10$. To underline the difference
between the stationary and dynamical clustered states at $\tau=5$ and $\tau=10$, we plot side-by-side the corresponding pairwise coherence matrices $\left\{\kappa_{ij}\right\}$, see Figs. \ref{Fig3}(a) and \ref{Fig3}(b), where the network nodes have been rearranged by a hierarchical clustering algorithm according to a
form of metric distance that has the most coherent nodes the closest. This makes it explicit how the
intercluster coherence for the $2$-cluster state is virtually negligible with respect to the $3$-cluster
case. Loosely speaking, within an unstable three-part population division, when a certain
fraction is firing, the other is refractory and the neurons in the smallest cluster are at
rest (excitable). This less clear separation is also apparent when comparing the nodal degree
distributions in cases $\tau=5$ and $\tau=10$, obtained if one assumes $\left\{\kappa_{ij}\right\}$
to provide weights for the network whose links stand for the correlated dynamics between the neurons.
For $\tau=5$, the bimodal degree distribution is clearly seen without raising the connectivity threshold,
whereas for $\tau=10$ the initially smeared three-modal distribution refines after some
thresholding is performed. The rationale of dynamical clustering may best be understood by analyzing the $r_i$
distribution in the $3$-cluster state. Apart from being wider than in the $2$-cluster state,
it peaks at a much smaller value $<r>_m\approx0.09$, implying the more regular neuron firing.
For this to hold, synchrony within the clusters has to be of intermittent nature, such that
the neurons once engaged in synchronized spiking are much more likely to do so again.

\begin{figure}[t]%2
%\begin{minipage}[t]{\columnwidth}
\centering
\includegraphics[scale=0.30]{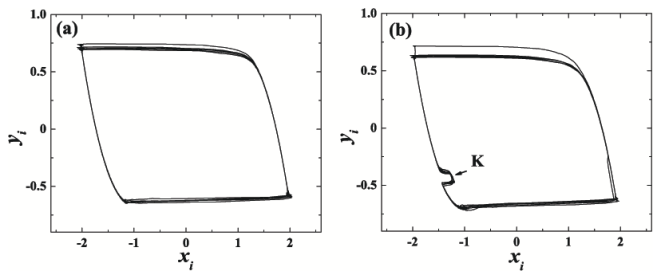}
%\end{minipage}
\caption{ (a) and (b) show typical PPs of neurons participating the homogeneous global oscillations
and clustered states, respectively. The latter are distinguished by a kink $K$, which is a
signature of the $D-\tau$ co-effect. The parameter sets are $\tau=6, D=0.0005, c=0.1$ in (a) and
$\tau=2, D=0.00025, c=0.1$ in (b). \label{Fig4}}
\vspace{-0.75cm}
\end{figure}

Aa understanding of clustering mechanism is revealed by comparing
the typical PPs of neurons participant in the homogeneous coherent state and
the clustered states, see Figs. \ref{Fig4}(a) and \ref{Fig4}(b). A striking feature in the
latter case is a kink at the refractory branch of the slow manifold. The appearance of a
kink is the key manifestation of the $D-\tau$ co-effect, that consists in separating
the ensemble into clusters and maintaining the proper phase difference between them. The purpose
of the kink is to keep the neurons frustrated long enough at the refractory branch before being
allowed to slide down to its left knee. This may be imagined as a form of a lock-and-release
behavior, where the delay primarily gives rise to the first, and noise to the second part. If a
fraction of the ensemble were to move beyond the left knee and the other were to lag behind, the
split should be amplified with each population cycle, eventually becoming resilient to
perturbation precisely due to trapping at the refractory branch. For trapping to be
successful, the kink has to be placed properly, approximately where the dynamics of the
representative point is most susceptible to perturbation along the slow manifold. Then, for a
brief period, due to an influence from $x_i$, the evolution of $y_i$ is locally accelerated,
becoming comparable to a speed of change in the direction orthogonal to the slow manifold, driven
by the spiking fraction of the population. Note that the trapping interval has to be adjusted so
that the entire population is entrained to a single frequency of firing. The latter matches
the one in delay-free case, which warrants stability against perturbations. The arguments
above and the numerical data seem to indicate how the delays where the coherence resonance is felt
the strongest may be approximated by $\tau=T_0/2+n*T_0$, with $T_0$ being the period of coherent
oscillations at $\tau=0$. Noise-wise, with increasing $\tau$, $D$ has to be adjusted to
higher values to regulate the relaxation from the kink to the slow manifold while maintaining
the entrainment to the proper frequency. In parallel, for stronger $D$, the representation cloud
of the firing fraction tends to disperse more, requiring a sufficient $\tau$ for this effect to be
averaged out.

\begin{figure}[h]%2
%\begin{minipage}[t]{\columnwidth}
\centering
\includegraphics[scale=0.31]{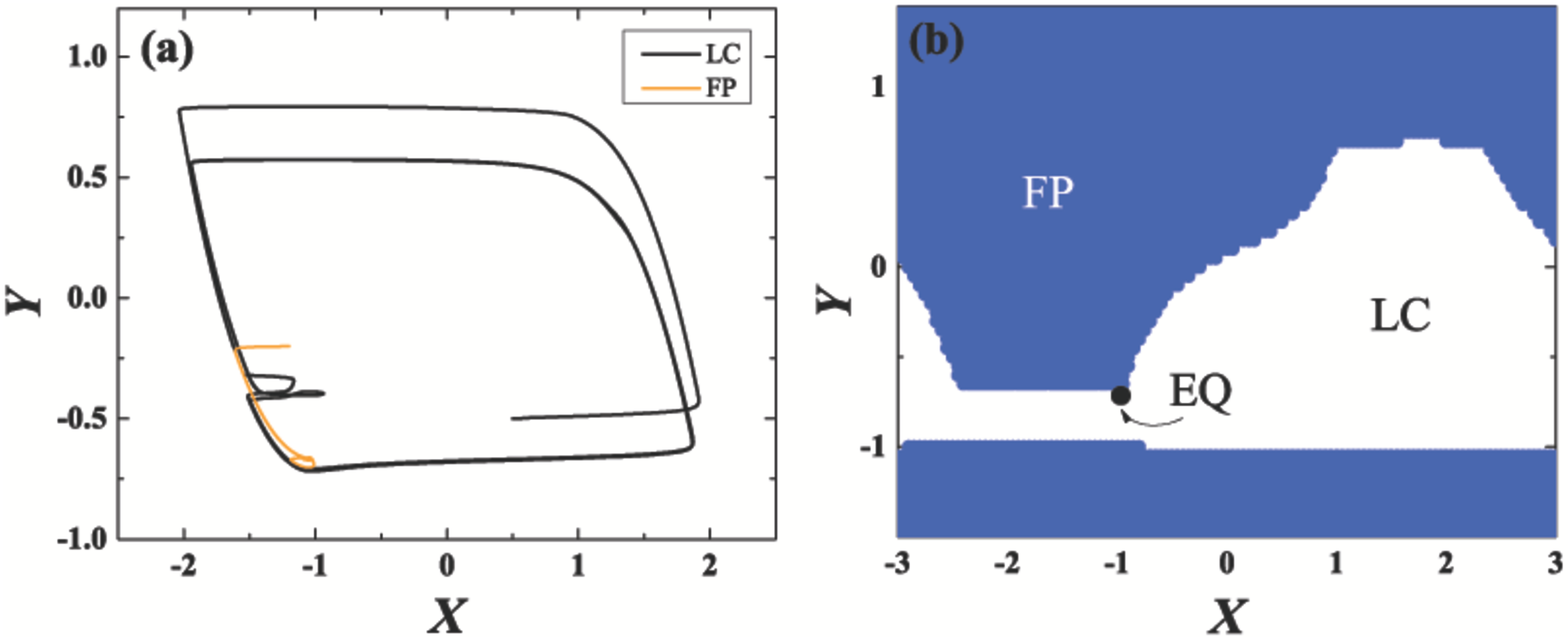}
%\end{minipage}
\caption{(color online). Bistability in the MF model: (a) shows the trajectories converging either
to the FP or the LC, depending on the initial conditions, whereas in (b) are displayed the two
basins of attraction for $\tau=2, D=0.00025, c=0.1$.
\label{Fig5}}
\vspace{-.8cm}
\end{figure}

The interplay between $D$ and $\tau$ is further highlighted by exploring the behavior of the
MF model (\ref{eq4}). Local bifurcation analysis shows that the MF exhibits a
succession of super- and subthreshold Hopf bifurcations \cite{us4}, which account for the transition from
the stochastically stable FP to the stable LC. Still, this scenario is confined to noise
higher than here: analytical and numerical means corroborate the Hopf bifurcations to
emerge about $D\approx0.0025$ at relevant $\tau$. Now we argue that the approximate model
is in qualitative terms able to capture the clustering effect occurring for small $D,c$ and
$\tau$. Focus is on the finding that MF system predicts an onset of cluster states by undergoing
a global bifurcation for the parameter values around $\tau=2, D=0.00025$ and $c=0.08$. At the
given $\tau$ and $D$, for $c<0.08$ the approximate model has only the equilibrium,
whereas around $c\simeq0.08$ a large and a small LC are born via a fold-cycle scenario.
Note how then the PP of the MF acquires the form qualitatively similar to those of the exact system's
in Figs. \ref{Fig2}(a) and \ref{Fig2}(b). The two sections of the emerging MF orbit mimic the action
of the fractions within the full population. This structure of the LC becomes unstable under
increasing $c$ or $\tau$, i.e. for the stronger impact of the interaction term. Another interesting
aspect to the approximate system is that it shows the complex LC to coexist with the FP, viz. the
basins of attraction in Fig. \ref{Fig5}(b), which is a feature apparently absent in the exact model.
However, the FP is located very close to the basins' boundary which indicates it to be stochastically
unstable in the exact system for an arbitrary small noise.

%Though the MF approximation cannot be expected to hold for the higher $D$ and $\tau$, one
%can still gain from it some insight into the behavior of the exact system. With the stochastic effects
%abstracted in Eq. (\ref{eq4}), the increase of $\tau$ leaves the orbit to stick in the vicinity of the
%FP for too long, which distorts the LC as it misses out on the proper entrainment frequency. This
%discrepancy can prove potentially revealing, as it suggests how the role of noise for the resonance effect
%may lie in maintaining the distance of the kink from the separatrix between the FP and LC.

We have reported on a novel phenomenon where clustering within the homogeneous
neural population is induced by an interplay of noise and time delay. This paradigm is
distinct from most current explanations on how the clustered states may arise, for it
does not treat $D$ and $\tau$ as destabilizing and detrimental, but rather as biased toward
the formation of dynamical structure in networks that are unstructured both in terms of
topology and local parameters. The analyzed model is minimal yet sufficient to display
an interesting type of behavior, possible only as an interplay of excitability, noise and
interaction delay. Once the phenomenon is recognized as caused only by these qualitative
properties one can study the effects of more realistic assumptions on the distribution of
neuronal properties and connection patterns. 
%Another interesting point is that the behavior
%of the derived MF model can aid in understanding the precise roles played by $D$ and $\tau$.
An interesting point concerns the derived MF model, which can aid in understanding 
the precise roles played by $D$ and $\tau$. Notably, beneath the surface lies a more stratified 
phenomenon, where the subtle adjustment between the parameters affects the number of clusters, 
their configuration, stationary or dynamical character, as well as whether the cluster states 
occur monostable or coexist with the homogenous solution at the given population size. This 
framework could find application within the research on neural systems and other excitable media.

\begin{acknowledgments}
\vspace{-0.15cm}
This work was supported in part by the Ministry of Education and
Science of the Republic of Serbia, under project Nos. $171017$ and
$171015$.
\vspace{-.85cm}
\end{acknowledgments}

\end{document}